\documentclass[twocolumn]{revtex4}
\usepackage{graphicx}
\begin{document}
%
%

\title{Exchange Interaction and $T_c$ in Alkaline-earth-metal-oxide-based DMS without Magnetic Impurities: First Principle Pseudo-SIC and Monte Carlo Calculation}
\author{Van An Dinh}
\email{divan@cmp.sanken.osaka-u.ac.jp}
\author{Masayuki Toyoda, Kazunori Sato}
\author{Hiroshi Katayama-Yoshida}
\affiliation{The Institute of Scientific and Industrial Research, Osaka University, 8-1 Mihogaoka, Ibaraki, Osaka 567-0047, Japan.}
\begin{abstract}
 The prospects of half-metallic ferromagnetism being induced by the incorporation of C atoms into alkaline-earth-metal-oxides are investigated by the first principle calculation. The origin of the ferromagnetism is discussed through the calculation of the electronic structure and exchange coupling constant by using the pseudo-potential-like self-interaction-corrected local spin density method. The Curie temperature ($T_c$) is also predicted by employing the Monte Carlo simulation. It is shown that by taking the electron self-interaction into account, the half-metallic ferromagnetism induced by C in the host materials is more stabilized in comparison with the standard LDA case, and the C's $2p$ electron states in the bandgap become more localized resulting in the predominance of the short-ranged exchange interaction. While the ferromagnetism in MgO$_{1-x}$C$_x$ is stabilized due to the exchange interaction of the $1st$-nearest neighbor pairs and might be suppressed by the anti-ferromagnetic super-exchange interaction at higher $x$, the ferromagnetism in CaO$_{1-x}$C$_x$, SrO$_{1-x}$C$_x$, and BaO$_{1-x}$C$_x$ is stabilized by both the $1st$- and $2nd$-nearest neighbor pairs, and $T_c$ monotonously increases with the C concentration.   
\end{abstract}
%
\maketitle

Besides the attempts to discover dilute magnetic semiconductors (DMSs) by incorporating transition metals into various materials to realize ferromagnetic DMSs for spintronic devices, recently the ferromagnetism induced by nonmagnetic impurities has  also attracted great attention among both theoretical and experimental scientists. A novel class of magnetic materials can be formed by incorporating nonmagnetic impurities or by lattice defects. With regard to this issue, the magnetism induced by the cation vacancy in MgO \cite{Halliburton}, Ca vacancy in CaO \cite{elfimov,ken1}, Hf vacancy in HfO$_2$ with $T_c$ exceeding $500$ K\cite{Ven, Coey, Pemmaraji,Osorio}, and by hydrogen in graphite \cite{Kusabe,Esquinazi} and in carbon nanotubes \cite{Ma} etc. has been reported. It has also been predicted that magnetism can be induced by nonmagnetic impurities such as C and N by substituting the O atoms in various oxides. For example, the ferromagnetism caused by N and C in alkaline-earth-metal-oxides \cite{ken1, ken2} and the half-metallic ferromagnetism induced by N in quartz-SiO$_2$ \cite{An1}. 

It is suggested that the magnetism might arise in the host materials if impurities have a finite local magnetic moment that interacts with each other to form a magnetic moment net. In the materials in which the magnetism is induced by incorporating nonmagnetic impurities, the substitutional ions may have a nonzero magnetic moment and the $2p$-electrons of these ions, rather than the $3d$-electrons, play an essential role in inducing the magnetism in the host materials. They form an impurity band in the deep  bandgap, and ferromagnetism can be induced if the Fermi level lies in these impurity bands. Furthermore, recent research also shows the role of the $2p$-like impurity band formed by C and N in the stabilization of the ferromagnetism in  Ga$_{1-x}$Mn$_x$As \cite{An21} and In$_{1-x}$Mn${_x}$N \cite{An22}.
 
 In this Letter, we will discuss the origin of the ferromagnetism that is induced by substituting C for O in four alkaline-earth-metal-oxides $-$ MgO, CaO, BaO, and SrO $-$ through the calculation of not only the electronic structure but also the exchange coupling constant and Curie temperature ($T_c$). It should be noted that the prediction of ferromagnetism and $T_c$ in ref. 3 and ref. 11 is based on the electronic structure calculation using the density functional theory within the local density approximation (LDA) and the mean field approximation (MFA) which cannot be successful for describing many materials, especially for strongly correlated systems. The LDA often overestimates the hybridization between electron states due to the underestimation of the bandgap energies of semiconductors. Moreover, the calculation of $T_c$ within the MFA often predicts $T_c$ with an excessively high value even if  the substitutional concentration is lower than the percolation limit \cite{sato1}. Therefore, in order to describe and predict the properties of these systems, we need a more accurate method. There have been several attempts to improve the LDA to overcome these inaccuracies. One of the most popular methods is LDA+U which improves the LDA by taking the strong correlation effects into account through the screened Coulomb parameter $U$ \cite{Asimov}. The other popular approaches are the self-interaction correction (SIC) \cite{Svane} and the recent pseudo-SIC \cite{Filippetti} methods. SIC methods differ from LDA+U in that LDA+U uses the additional parameter $U$, while the SIC methods do not require any additional parameter at all. Consequently, in order to investigate the origin of ferromagnetism in alkaline-earth-metal-oxide-based DMSs, we apply the pseudo-SIC method based on the improvement of the MACHIKANEYAMA2000 package coded by Akai \cite{Akai}  (LDA+SIC) in the calculation of the electronic structures, and then we use the formula by Liechtenstein {\it et al}.\cite{Liechtenstein} to calculate the exchange interaction $J_{ij}$ between two impurities at the $i$th- and $j$th-sites in the ferromagnetic coherent potential approximation medium. Finally, we employ the Monte Carlo simulation to estimate $T_c$.
 
  The substitution of O with C in alkaline-earth-metal-oxides is treated randomly. For convenience, the lattice constants of alkaline-earth-metal-oxide-based DMSs are fixed to the values of undoped crystals \cite{Wyckoff} ($a=4.123$ \AA, $4.909$ \AA, $5.160$ \AA, and $5.520$ \AA\ for MgO, CaO, BaO, and SrO, respectively) and no distortion in lattice structure is assumed. Throughout the electronic structure calculations, 624 independent $k$-sampling points in the first Brillouin zone are used. The potential form is restricted to the muffin-tin type, and  muffin-tin radii are chosen in such a way that the ions at lattice sites can touch each other. The relativistic effect is also taken into account by using scalar relativistic approximation.

\begin{figure}
\begin{center}\leavevmode
\includegraphics[width=1.0\linewidth]{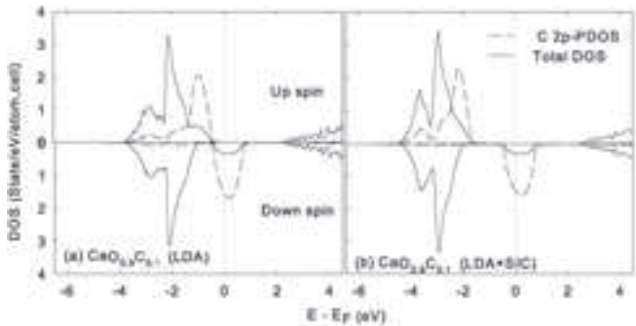} 
\caption{ Total DOS (solid line) and C's $2p$ - PDOS (dashed line) for CaO$_{0.9}$C$_{0.1}$. The left-hand figure (a) shows the standard LDA calculation and the other (b) shows the pseudo-SIC results. The upper plane corresponds to the majority spin and lower plane to the minority spin.}\label{f1}
\end{center}
\end{figure}

 In order to compare the density of states (DOS) calculated within the standard LDA with that within LDA+SIC we plot the DOS of the typical case of alkaline-earth-metal-oxides, CaO$_{1-x}$C$_x$, at $x=0.10$ in Fig.~\ref{f1}. The figure on the left-hand side corresponds to the DOS calculated within the standard LDA (Fig.~\ref{f1}(a)), and the figure on the right-hand side illustrates the DOS within LDA+SIC (Fig.~\ref{f1}(b)). It can be seen that, taking the self-interaction of electrons into account, the LDA+SIC calculation gives a wider bandgap than the standard LDA. While the position of the minority spin states of C's $2p$ electrons remains unchanged, the majority spin states calculated within LDA+SIC shifts about $1.1$ eV in comparison with the standard LDA one, leading to a higher localization of LDA+SIC $2p$ states than LDA. And the local magnetic moment of C increases from $1.242\mu_B$ (LDA) to $1.482\mu_B$ (LDA+SIC). The exchange splitting in the case of LDA+SIC increases approximately 2 times as compared to that in the case of LDA, resulting in the possibility of the suppression of super-exchange interaction and the enhancement of the ferromagnetic double exchange mechanism.
 \begin{figure}
\begin{center}\leavevmode
\includegraphics[width=1.0\linewidth]{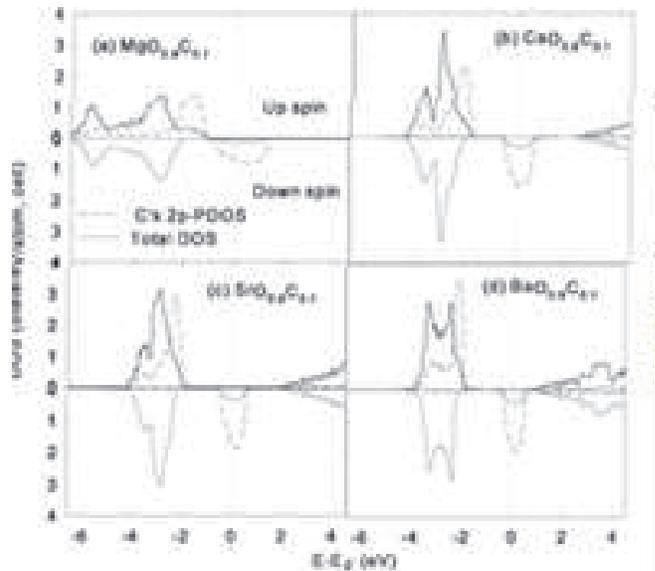}
\caption{ Total DOS (solid line) and C's $2p$ - PDOS (dashed line) obtained by pseudo-SIC calculation for  MgO$_{0.9}$C$_{0.1}$ (a), CaO$_{0.9}$C$_{0.1}$ (b), SrO$_{0.9}$C$_{0.1}$ (c), and BaO$_{0.9}$C$_{0.1}$ (d).}\label{f2}
\end{center}
\end{figure}

\begin{figure}
\begin{center}\leavevmode 
\includegraphics[width=0.90\linewidth]{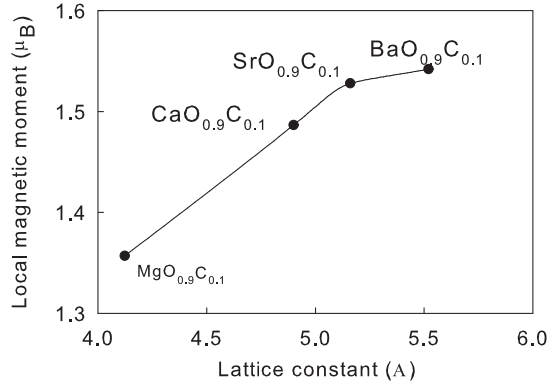}
\caption{Local magnetic moment of substitutional C atoms in MgO$_{1-x}$C$_{x}$, CaO$_{1-x}$C$_{x}$, SrO$_{1-x}$C$_{x}$, and BaO$_{1-x}$C$_{x}$ at $x$~$=$~$0.10$.}\label{f3}
\end{center}
\end{figure}

\begin{figure} 
\begin{center}
\includegraphics[width=0.90\linewidth]{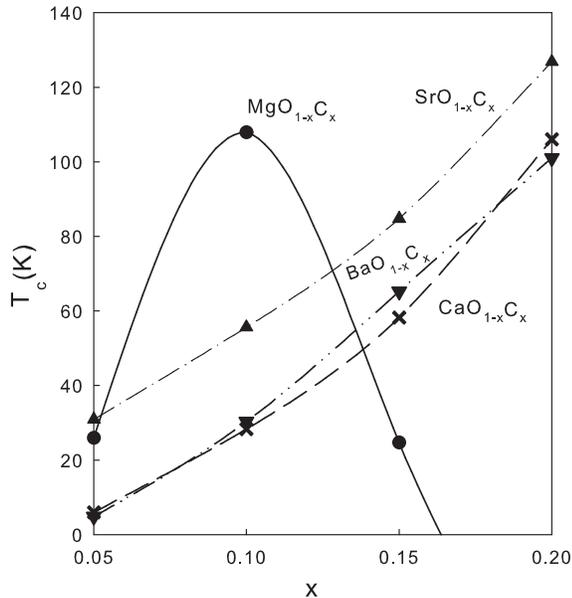}  
\caption{Curie temperature $T_c$ vs. substitutional impurity concentration  in MgO$_{1-x}$C$_{x}$ (solid line), CaO$_{1-x}$C$_{x}$ (dashed line), SrO$_{1-x}$C$_{x}$ (dashed-dot-dot line), and BaO$_{1-x}$C$_{x}$ (dashed-dot line). The circles, crosses, and triangles denote the calculated points.}\label{f5}
\end{center} 
 \end{figure}

Figure~\ref{f2} depicts the DOS of AO$_{1-x}$C$_x$ (A = Mg, Ca, Sr, Ba) calculated within LDA+SIC at $x=0.10$.  As seen from Fig.~\ref{f2}, all materials in question are half-metallic. However, the exchange mechanism causing the ferromagnetism is somewhat different. Figure~\ref{f2}(a) illustrates the DOS of MgO$_{0.9}$C$_{0.1}$. With the smallest lattice constant, MgO$_{0.9}$C$_{0.1}$ has the largest bandgap energy. The C's $2p$ states are located near the top of the valence band originated from anion $p$-states, resulting in the strong hybridization of $2p$ electron wave functions. The majority spin states hybridize with O's $2p$ states, leading to the appearance of the impurity band that connects to the top of the valence band and causes a narrower majority spin bandgap. The minority spins create an impurity band in the bandgap, which includes the Fermi level. This impurity band is broadened with a half-width of about $1.5$ eV. The exchange splitting in this material approximates to $2.14$ eV and is the smallest compared with the three remaining materials. These impurity bands can be broadened more strongly with increasing C concentration, leading to the antiferromagnetic super-exchange interaction being easy to occur and the compensation of majority and minority spins. These mean that the ferromagnetism might be suppressed at higher C concentrations. In addition, our calculations for higher C concentrations show that the ferromagnetism in MgO$_{1-x}$C$_{x}$ is most stabilized at $x\approx 0.10$ (also refer to Fig.~\ref{f5}). Figures~\ref{f2}(b)$-$\ref{f2}(d) demonstrate the DOS of CaO$_{0.9}$C$_{0.1}$, SrO$_{0.9}$C$_{0.1}$, and BaO$_{0.9}$C$_{0.1}$, respectively. As seen from the figures, the bandgap becomes narrower and the localization of the $2p$ states becomes stronger with increasing the distance between atoms (or lattice constant). The majority spin states are located in the valence band and cause a small broadening of the band. The Fermi level lies in the impurity band induced by minority spins with 1/3 of this impurity band being occupied by electrons. The exchange splitting in these materials is approximately $2.3$ eV, but the C's $2p$ states in  BaO$_{0.9}$C$_{0.1}$ is the most localized. This change in the localization with respect to the lattice constant is also indicated by the value of the local magnetic moment of C which is shown in Fig.~\ref{f3}. As a result, one can expect the ferromagnetism to be stabilized by the predominant double exchange mechanism in these materials even at higher C concentrations.

\begin{figure}
\begin{center}\leavevmode 
\includegraphics[width=1.0\linewidth]{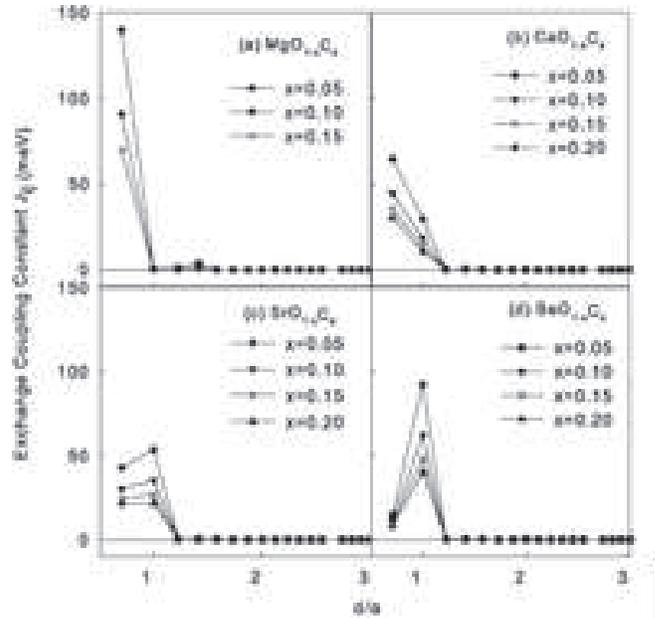}
\caption{Exchange coupling constant  vs. distance between C atoms in units of the lattice constant $a$ in MgO$_{1-x}$C$_{x}$ (a), CaO$_{1-x}$C$_{x}$ (b), SrO$_{1-x}$C$_{x}$ (c), and BaO$_{1-x}$C$_{x}$ (d) at several concentrations of C ($x = 0.05, 0.10, 0.15$, and $0.20$).}\label{f4}
\end{center}
\end{figure}

The exchange interactions $J_{ij}$ in MgO$_{1-x}$C$_{x}$, CaO$_{1-x}$C$_{x}$, SrO$_{1-x}$C$_{x}$, and BaO$_{1-x}$C$_{x}$ are shown in Fig.~\ref{f4}. As seen from Fig.~\ref{f4}(a), the exchange interaction of the $1st$-nearest neighbors in MgO$_{1-x}$C$_{x}$ is the strongest. The interaction strength is considerably strong (about $140$ meV for 5\%
 of the C concentration) and most of the contributions come from the nearest neighbor interaction $J_{01}$. Except $J_{01}$ and $J_{04}$ (at 10\%
  of the C concentration), contributions from other $J_{ij}$ are very small and ignorable. Thus, the exchange interaction in MgO$_{1-x}$C$_{x}$ can be considered as a typical case of short ranged interactions. In addition, the exchange interaction is considerably suppressed as the C concentration increases. At $x=0.20$, the exchange interaction becomes very small and the magnetism is fully suppressed. Interestingly, by increasing the lattice constant, the role of the $2nd$-nearest neighbors ($J_{02}$) becomes more important. For CaO$_{1-x}$C$_{x}$ (Fig.~\ref{f4}(b)), $J_{01}$ is approximately two times smaller than that in MgO$_{1-x}$C$_{x}$, but the contributions $J_{02}$ from the $2nd$-nearest neighbors becomes important. The dominance of the contributions $J_{02}$ of the $2nd$-nearest neighbor pairs in SrO$_{1-x}$C$_{x}$ and BaO$_{1-x}$C$_{x}$ is shown in Fig.~\ref{f4}(c) and Fig.~\ref{f4}(d), respectively. In these former materials, the exchange interactions of the $2nd$-nearest neighbors are more important than the $1st$-nearest neighbors. This can be caused by the sufficiently large distances between the $1st$-nearest neighbors, while the interactions between the 2$nd-$nearest atoms easily accur through the mediate atoms such as Sr or Ba which have large ionic radii. 

In short, the role of the nearest neighbors in the exchange interaction changes with the lattice constant. The higher the lattice constant, the greater is the contribution from the $2nd$-nearest neighbors in the induction of the magnetism in the host materials. The exchange interaction is very short ranged and the ferromagnetic double exchange mechanism is expected to be predominant for all materials in question. Except MgO$_{1-x}$C$_{x}$, the magnetism in CaO$_{1-x}$C$_{x}$, SrO$_{1-x}$C$_{x}$, and BaO$_{1-x}$C$_{x}$ can be more stabilized at higher C concentrations. This is also consistent with the results of the Monte Carlo simulation for the estimation of $T_c$.

  In order to evaluate $T_c$, we perform the Monte Carlo simulations. The Metropolis algorithm \cite{Metropolis} is applied to calculate the thermal average of the magnetization $M$ and its powers. Then, the cumulant crossing method proposed by Binder \cite{Metropolis} is employed and the fourth order cumulant $U_4$ (a linear combination of $<\!\!M^4\!\!>\!\!\!/\!\!\!<\!\!M^2\!\!>^2$) is calculated as a function of temperature for different cell sizes ($14\times 14\times 14, 16\times 16\times 16$, and $18\times 18\times 18$ conventional fcc cells) to find the universal fix-point at $T_c$. We estimate $T_c$ for four values of the substitutional C concentrations: $x=0.05, 0.10, 0.15$, and $0.20$. The obtained results are demonstrated in Fig.~\ref{f5}. Some points should be clarified here.
  
First, as discussed above for Fig.~\ref{f2}(a) and Fig~\ref{f4}(a), the ferromagnetism of MgO$_{1-x}$C$_{x}$ is the most stabilized at $x\approx 0.10$ and will be suppressed at higher C concentrations due to the anti-ferromagnetic super-exchange interaction and the suppression of the majority and minority spins. $T_c$ of MgO$_{1-x}$C$_{x}$  (solid line) increases with $x$ in the range of $x$ from 0.05 to $0.10$ and has a peak of $108$~K at $x\approx 0.10$, and then it sharply drops as a function of $-x$ at higher $x$. As shown in Fig.~\ref{f2}(a), the impurity band formed by C's $2p$ electrons is strongly broadened. The bandwidth $W$ of the impurity band in the gap can be much larger than the effective correlation energy $U= E(N+1)+E(N-1)-2E(N)$, and the Stoner's condition for the existence of magnetism might be contravened \cite{ken2, An1}. Correspondingly, the magnetism might be fully suppressed at the C concentrations higher than $16\%$
 because of the strong broadening of the impurity bands of both majority and minority spins, resulting in the compensation of spins; therefore, the material in question has a nonmagnetic behavior at higher C concentrations.
 
   Second, contrary to MgO$_{1-x}$C$_{x}$, the impurity states in CaO$_{1-x}$C$_{x}$, SrO$_{1-x}$C$_{x}$, and BaO$_{1-x}$C$_{x}$  are localized and the bandwidths of the impurity bands formed by minority spins in the gap are sufficiently small to satisfy Stoner's condition; hence,  the ferromagnetism  is stabilized by a predominant ferromagnetic double exchange mechanism and is more stabilized as $x$ increases. Although the exchange interaction $J_{01}$ between the $1st$-nearest neighbor pairs considerably decreases with the increasing lattice constant, owing to the contributions $J_{02}$ from the $2nd$-nearest neighbors, $T_c$ monotonously increases with $x$ and can gain a value higher than room temperature if the C concentration is high enough. However, a question that arises here is the solubility of C in the materials. 
  
In conclusion, we have presented the results of the study on the origin of ferromagnetism and predicted $T_c$ of alkaline-earth-metals-oxide-based DMSs without transition metal elements. The electronic structures and exchange coupling constants are calculated by applying the pseudo-SIC approach.  The dominant exchange mechanism and the role of the nearest neighbors in these DMSs are discussed, and $T_c$ is also evaluated by employing the Monte Carlo simulation. In short, some comments can be made as follows:
  
  ({\it i}) All DMSs in question have a half-metallic ferromagnetism. The ferromagnetic double exchange mechanism is predominant for all materials in question, except for MgO$_{1-x}$C$_{x}$ when the C concentrations is higher than 10\%.
  
  ({\it ii}) The exchange interaction in these materials is strong but short ranged. While the contributions come mostly from the $1st$-nearest neighbor pairs in MgO$_{1-x}$C$_{x}$, the important role played by the $1st$- and $2nd$-nearest neighbor pairs in CaO$_{1-x}$C$_{x}$, SrO$_{1-x}$C$_{x}$, and BaO$_{1-x}$C$_{x}$ is comparable. However, while $J_{01}$ becomes weaker, $J_{02}$ gets considerably stronger with the lattice constant being larger and becomes the dominant contribution to the stabilization of the ferromagnetism in SrO$_{1-x}$C$_{x}$ and BaO$_{1-x}$C$_{x}$. 
 
  ({\it iii}) Correspondingly, $T_c$ of CaO$_{1-x}$C$_{x}$, SrO$_{1-x}$C$_{x}$, and BaO$_{1-x}$C$_{x}$  increases monotonously with the increasing C concentration, while $T_c$ of MgO$_{1-x}$C$_{x}$ gains the highest value at 10\%
   of the C concentration, then it drops sharply and tends to reach zero at higher concentrations due to the anti-ferromagnetic superexchange interaction and the compensation of majority and minority spins.
\begin{acknowledgments}
 This research was partially supported by a Grant-in-Aid for Scientific Research in Priority Areas ``Quantum Simulators and Quantum Design" (No. 17064014) and ``Semiconductor Nanospintronics," a Grand-in-Aid for Scientific Research for young researchers, JST-CREST, NEDO-nanotech, the 21st Century COE, and the JSPS core-to-core program ``Computational Nano-materials Design." We are grateful to Prof. H. Akai (Osaka Univ.) for providing us with the MACHIKANEYAMA2000 package and to Prof. A. Yanase (Osaka Univ.) for many valuable discussions.
\end{acknowledgments}

 \end{document}